\documentclass[12pt,preprint]{article}%
\usepackage{amssymb}
\usepackage{amsmath}
\usepackage{graphics}
\usepackage{epsfig}
\usepackage{amsfonts}
\usepackage{graphicx}%
\renewcommand{\vec}[1]{{\bf #1}}
\newcommand{\eqb}{\begin{equation}}
\newcommand{\eqe}{\end{equation}}
\newcommand{\dmb}{\begin{displaymath}}
\newcommand{\dme}{\end{displaymath}}
\newcommand{\pd}{\partial}

\newcommand{\eab}{\begin{eqnarray}}
\newcommand{\eae}{\end{eqnarray}}

\newcommand{\be}{\begin{equation}}
\newcommand{\ee}{\end{equation}}

\begin{document}
\begin{titlepage}
\begin{flushright}
\end{flushright}
\vspace{0.6cm}
\begin{center}
\Large{Thermal ground state and nonthermal probes}  
\vspace{0.5cm}\\ 
\small{Thierry Grandou$^\dagger$ and Ralf Hofmann$^{*}$}
\end{center}
\vspace{0.5cm}
\begin{center}
{\em $\mbox{}^\dagger$ Institut Nonlin\'{e}aire de Nice\\ 
1361 route des Lucioles, Sophia Antipolis\\  
F-06560 Valbonne, France}
\end{center}
\begin{center}
{\em $\mbox{}^*$ Institut f\"ur Theoretische Physik\\
Universit\"at Heidelberg\\
Philosophenweg 16\\
69120 Heidelberg, Germany}
\end{center}
\vspace{1.5cm}

\begin{abstract}

The Euclidean formulation of SU(2) Yang-Mills thermodynamics admits periodic, (anti)selfdual  
solutions to the fundamental, classical equation of motion which possess  
one unit of topological charge: (anti)calorons. A spatial coarse graining over the central region in a pair of such localised 
field configurations with trivial holonomy generates an 
inert adjoint scalar field $\phi$, effectively describing the pure quantum 
part of the thermal ground state in the induced quantum field theory. 
Here we show for the limit of zero holonomy how (anti)calorons associate a temperature independent 
electric permittivity and magnetic permeability to the thermal ground 
state of SU(2)$_{\tiny\mbox{CMB}}$, the Yang-Mills theory 
conjectured to underlie the fundamental description of thermal photon gases. 

\end{abstract}
\vspace{0.5cm}
$\mbox{}^*$email: r.hofmann@thphys.uni-heidelberg.de

\end{titlepage}

\section{Introduction}

Quantum Mechanics is a highly efficient framework to describe the subatomic 
world \cite{Heisenberg, Dirac, Feynman}, including coherence phenomena that extend to macroscopic length 
and time scales \cite{BCS,Josephson,Abrikosov}. The key quantity to describe deviations from classical behavior is 
Planck's quantum of action $\hbar=\frac{h}{2\pi}=6.58\times 10^{−16}\,$eV\,s which determines the fundamental interaction between charged matter and the electromagnetic field and thus also the shape of 
blackbody spectra by relating frequency $\omega$ and wave vector $\vec{k}$ 
to particle-like energy $E=\hbar\omega$ and momentum $\vec{p}=\hbar\vec{k}$ and by appeal to 
Bose-Einstein statistics. In Quantum Mechanics, $\hbar$ sets the strength of 
multiplicative noncommutativity for a pair of canonically conjugate variables such as 
position and momentum, implying the respective uncertainty relations.  

Although generally accepted as a universal constant of nature and in spite 
of the fact that we are able to efficiently compute quantum mechanical amplitudes 
and quantum statistical averages for a 
vast variety of processes in particle collisions, atoms and molecules, 
extended condensed-matter systems, and astrophysical objects to match  
experiment and observation very well, one should remain 
curious concerning the principle mechanism that causes the 
emergence of a universal quantum of action. In \cite{KavianiHofmann2012,KrasowskiHofmann2014} it 
was argued that the irreconcilability of classical Euclidean and Minkowskian 
time evolution as expressed by a time-periodic SU(2) (anti)selfdual gauge field 
configuration -- a (anti)caloron --, whose action $\hbar$ is 
associated with one unit of winding about a central spacetime point, gives rise to 
indeterminism in the process it mediates. That each unit of action assigned to (anti)calorons of 
radius $\rho=|\phi|^{-1}$, which dominate the emergence of the thermal ground state, equals 
$\hbar$ follows from the value of the coupling $e$ in 
the induced, effective, thermal quantum field theory \cite{loopExp} of 
the deconfining phase in SU(2) Yang-Mills thermodynamics. The coupling 
$e$, in turn, obeys an  
evolution in temperature (flat almost everywhere) which represents the validity of 
Legendre transformations in the effective ensemble where the thermal 
ground state co-exists with massive (adjoint Higgs mechanism) and 
massless (intact U(1)) thermal fluctuations. 
The thermal ground state thus is a spatially homogeneous 
ensemble of quantum fluctuations carried by (anti)caloron centers. 
At the same time, as we shall see, this state provides electric and magnetic 
dipole densities supporting the propagation of electromagnetic waves in an SU(2) Yang-Mills of 
scale $\Lambda\sim 10^{-4}\,$eV, SU(2)$_{\tiny\mbox{CMB}}$ \cite{Hofmann2012}.             

In the present work, we establish this link between quantised 
action, represented by $\phi$, and classical wave propagation enabled by  
the vacuum parameters $\epsilon_0$ and $\mu_0$ in terms of the central and 
peripheral structure of a trivial-holonomy (anti)caloron, respectively. That is, by allowing a 
fictitious temperature $T$ to represent the energy density of an electromagnetic wave (nonthermal, external probe) via the 
thermal ground state through which it propagates we ask what this implies for $\epsilon_0$ and $\mu_0$. 
As a result, both $\epsilon_0$ and $\mu_0$ neither depend on $T$ nor, as we shall argue, on any 
singled-out inertial frame. But this means no more and no less than the rivival of 
the luminiferous aether, albeit now in a Poincar\'{e} invariant way.  

This paper is organised as follows. In the next section we shorty 
discuss key features of the effective theory for the deconfining phase 
of SU(2) Yang-Mills thermodynamics. Sec.\,\ref{CS} contains 
a reminder to principles in interpreting a 
Euclidean field configuration in terms of Minkowskian observables. In a next step, general facts are reviewed 
on Euclidean, periodic,(anti)selfdual field configurations of charge 
modulus unity concerning the central locus of action, their holonomy, and 
their behaviour under semiclassical deformation. 
Finally, we review the anatomy of zero-holonomy Harrington-Shepard (HS) 
caloron in detail, pointing out its staticity for 
spatial distances from the center that exceed the inverse of temperature, 
and discuss which static charge configuration it resembles depending 
on two distinct distance regimes. In 
Sec.\,\ref{EMvacTGS} we briefly review the postulate 
that an SU(2) Yang-Mills theory of scale 
$\Lambda\sim 10^{-4}\,$eV, SU(2)$_{\tiny\mbox{CMB}}$, 
describes thermal photon gases \cite{Hofmann2012}. 
Subsequently, the large-distance regime 
in a HS (anti)caloron is considered in order to deduce an expression for 
$\epsilon_0$ based on knowledge about the electric dipole moment provided by a (anti)caloron of 
radius radius $\rho=|\phi|^{-1}$, the size of the spatial coarse-graining volume $V_{\tiny\mbox{cg}}$, and the fact 
that the energy density of the probe must match that of 
the thermal ground state. As a result, $\epsilon_0$ and $\mu_0$ 
turn out to be $T$ independent, the former representing 
an electric charge, large on the scale of the electron charge, of the fictitious constituent monopoles giving rise 
to the associated dipole density. Zooming in to smaller spatial distances 
to the center, the HS (anti)caloron exhibits isolated (anti)selfdual monopoles. For them to turn into dipoles shaking by the 
probe fields is required. We then show that the definitions of $\epsilon_0$ and 
$\mu_0$, which were successfully applied to the large-distance regime, become 
meaningless. Finally, our results are discussed. 
Sec.\,\ref{SC} summarises the paper and discusses the universality of 
$\epsilon_0$ and $\mu_0$ for the entire electromagnetic spectrum.

\section{Sketch of deconfining SU(2) Yang-Mills thermodynamics}

For deconfining SU(2) Yang-Mills thermodynamics, a spatial 
coarse graining over the (anti)selfdual, 
that is, the nonpropagating \cite{Shuryak}, 
topological sector with charge modulus $|Q|=1$ can be performed, see \cite{Hofmann2012} and references therein, 
to yield an inert adjoint scalar 
field $\phi$. Its modulus $|\phi|$ sets the maximal possible resolution in the effective theory whose ground state energy 
density essentially is given as $\mbox{tr}\,\frac{\Lambda^6}{\phi^2}=4\pi\Lambda^3 T$ 
($\Lambda$ a constant of integration of dimension mass) and whose propagating sector 
is, in a totally fixed, physical gauge (unitary-Coulomb) characterised by a 
massless mode ($\gamma$, unbroken U(1) subgroup of SU(2)) and two thermal quasiparticle modes of 
equal mass $m=2e\,|\phi|$ ($V^\pm$, mass induced by adjoint Higgs mechanism) 
which propagate thermally, that is, on-shell only. Interactions within this propagating sector are mediated 
by isolated (anti)calorons whose action is argued to be $\hbar$ 
\cite{KavianiHofmann2012,KrasowskiHofmann2014}. Judged in terms of inclusive quantities 
such as radiative corrections to the one-loop pressure or the energy density of 
blackbody radiation, these interactions are feeble \cite{Hofmann2012}, and 
their expansion into 1-PI irreducible bubble diagrams is conjectured to terminate 
at a finite number of loops \cite{Hofmann2006}. However, spectrally seen, the effects 
of $V^\pm$ interacting with $\gamma$ lead to severe 
consequences at low frequencies and temperatures comparable to the 
critical temperature $T_c$ where screened (anti)monopoles, released by (anti)caloron dissociation upon 
large-holonomy deformations \cite{Diakonov2004}, rapidly become massless and thus 
start to condense.

\section{Caloron structure\label{CS}}

\subsection{Euclidean field theory and interpretable quantities\label{EMI}} 

Nontrivial solutions to an elliptic differential equation, such as the Euclidean Yang-Mills equation 
$D_\mu F_{\mu\nu}=0$, no longer are solutions of the corresponding hyperbolic 
equation upon analytic continuation $x_4\equiv\tau\to i x_0$ (Wick rotation). 
To endow meaning to quantities computed on classical field configurations on a 4D  
Euclidean spacetime in SU(2) Yang-Mills thermodynamics in terms of 
observables in a Minkowskian spacetime we thus must insist that these 
quantities are not affected by the Wick rotation. That is, to 
assign a real-world interpretation to a Euclidean quantity it needs to be 
(i) either stationary (not depend on $\tau$) or (ii) associated with an 
instant in Euclidean spacetime because, by exploiting time translational 
invariance of the Yang-Mills action, this instant can be picked as 
$(\tau=0,\vec{x})$ in Euclidean spacetime.

\subsection{Review of general facts\label{RGF}}

If not stated otherwise we work in supernatural units, $\hbar=c=k_B=1$, where $\hbar$ 
is the reduced quantum of action, $c$ the speed of light in vacuum, and 
$k_B$ Boltzmann's constant. A trivial-holonomy caloron of topological 
charge unity on the cylinder $S_1\times {\bf R}^3$, where $S_1$ is the circle of 
circumference $\beta\equiv 1/T$ ($T$ temperature) describing the 
compactified Euclidean time dimension ($0\le\tau\le\beta$), 
is constructed by an appropriate superposition of charge-one 
singular-gauge instanton prepotentials \cite{'tHooft1976} with the temporal 
coordinate of their instanton centers 
equidistantly stacked along the infinitely extended Euclidean time dimension 
\cite{HS1977} to enforce temporal periodicity, $A_\mu(\tau=0,\vec{x})=A_\mu(\tau=\beta,\vec{x})$. 
For gauge group SU(2) this Harrington-Shepard (HS) caloron is given as (antihermitian generators $t_a$ ($a=1,2,3$) with 
tr\,$t_at_b=-\frac12\delta_{ab}$):
\eqb
\label{HScaloron}
A_\mu=\bar{\eta}_{\mu\nu}^a t_a \pd_\nu\log\Pi(\tau,r)\,,
\eqe
where $r\equiv|\vec{x}|$, $\bar{\eta}_{\mu\nu}^a$ denotes the antiselfdual 
't Hooft symbol, $\bar{\eta}_{\mu\nu}^a=\epsilon^a_{\mu\nu}-\delta_{a\mu}\delta_{\nu 4}+
\delta_{a\nu}\delta_{\mu 4}$, and 
\eqb
\label{HSprepotential}
\Pi(\tau,r)=1+\frac{\pi\rho^2}{\beta r}\frac{\sinh\left(\frac{2\pi r}{\beta}\right)}{\cosh\left(\frac{2\pi r}{\beta}\right)-\cos\left(\frac{2\pi\tau}{\beta}\right)}\,.
\eqe 
Here $\rho$ is the scale parameter of the singular-gauge instanton to seed the 
"mirror sum" within $S^1\times {\bf R}^3$, leading to Eq.\,(\ref{HSprepotential}). The associated antiselfdual field configuration is obtained in replacing $\bar{\eta}_{\mu\nu}^a$ by $\eta_{\mu\nu}^a$ (selfdual 't Hooft symbol) 
in Eq.\,(\ref{HScaloron}). 

Configuration (\ref{HScaloron}) is singular at $\tau=r=0$. This point is the locus of the configuration's 
topological charge 
$Q=1$ in the sense that the integral of the Chern-Simons current $K_\mu=\frac{1}{16\pi^2}\epsilon_
{\mu\alpha\beta\gamma}\left(A^a_\alpha\pd_\beta A^a_\gamma+
\frac13\epsilon^{abc} A_\alpha^a A_\beta^b A_\gamma^c \right)$ over a three-sphere $S^3_\delta$ of 
radius $\delta$, which is centered there, 
yields unity independently of $\delta\ge 0$. Selfduality implies that the action of the 
HS caloron is given as
\eqb
\label{actHS}
S_C=\frac{8\pi^2}{g^2}\int_{S^3_\delta} d\Sigma_\mu K_\mu=\frac{8\pi^2}{g^2}\,,
\eqe
where $g$ is the coupling constant in Euclidean (classical) theory. Eq.\,(\ref{actHS}) holds 
in the limit $\delta\to 0$, meaning that $S_C$ can be attributed to the 
singularity of the HS solution at $\tau=r=0$ and thus has a Minkowskian intepretation, see Sec.\,\ref{EMI}. 
Based on \cite{loopExp} and on the fact 
that the thermal ground state emerges from $|Q|=1$ caloron/anticalorons, whose scale parameter $\rho$ essentially 
coincides with the inverse of maximal resolution, $|\phi|^{-1}$, in the 
effective theory for deconfining SU(2) Yang-Mills thermodynamics, it was argued 
in \cite{KavianiHofmann2012}, see also \cite{KrasowskiHofmann2014}, that $S_C$ (as well as the action of 
a HS anticaloron $S_A$ with $\rho\sim |\phi|^{-1}$) equals $\hbar$ if the effective theory is to be 
interpreted as a local quantum field theory.  

The HS caloron is the trivial-holonomy limit of the selfdual Lee-Lu-Kraan-van-Baal (LLKvB) 
configuration with $Q=1$ and total magnetic charge zero \cite{LeeLu1998,KraanVanBaal1998} 
which is constructed via the Nahm transformation of selfdual fields on the Euclidean four-torus \cite{Nahm1980}. For nontrivial holonomy ($A_4(r\to\infty)=iut^3$ 
with $0<u<\frac{2\pi}{\beta}$) the LLKvB solution exhibits a pair of a 
magnetic monopole (m) and its antimonopole (a) w.r.t. the Abelian subgroup U(1)$\subset$SU(2) left 
unbroken by $A_4(r\to\infty)\not=0$. Their masses are $m_m=4\pi u$ and 
$m_a=4\pi\left(\frac{2\pi}{\beta}-u\right)$ such that in the trivial-holonomy limits $u\to 0,\frac{2\pi}{\beta}$ one of these magnetic constituents becomes massless and thus completely spatially delocalised. 
For nontrivial holonomy, where both monopole and antimonopole are of finite mass, localised, and separated by a spatial distance
\eqb
\label{madistance}
s=\pi\frac{\rho^2}{\beta}\,,
\eqe  
they can be considered static by an exact cancellation 
of attraction, mediated by their U(1) magnetic fields, and repulsion due to the field $A_4$.   
As was shown in \cite{Diakonov2004} by investigating the effective action of a 
LLKvB caloron (integrating out Gaussian fluctuations), this balance is 
distorted, leading to monopole-antimonopole attraction 
for 
\eqb
0<u\le\frac{\pi}{\beta}\left(1-\frac{1}{\sqrt{3}}\right)\ \ \ \ \mbox{or}\ \ \ \ 
\frac{\pi}{\beta}\left(1+\frac{1}{\sqrt{3}}\right)<u\le\frac{2\pi}{\beta}\,, 
\eqe
and to repulsion in the complementary range of (large) holonomy. Because there is no localised counter part to a monopole or antimonopole in the trivial-holonomy limit, HS calorons must be considered 
stable under Gaussian fluctuations, in contrast to the case of nontrivial 
holonomy which is unstable. The latter statement is also mirrored by the fact that a nontrivial, static 
holonomy leads to zero quantum weight in the infinite-volume limit (which is 
realistic at high temperatures \cite{Hofmann2012} where the radius of the spatial coarse-graining volume for 
a single caloron diverges as $|\phi|^{-1}=\sqrt{\frac{2\pi T}{\Lambda^3}}$, $\Lambda$ 
the Yang-Mills scale). As a consequence, nontrivial holonomy can only occur transiently 
in configurations which do not saturate (anti)selfduality bounds to the Yang-Mills action. Again, this 
is equivalent to stating the instability of the LLKvB solution. It can be shown \cite{Hofmann2012} that the small-holonomy case of monopole-antimonople attraction by far dominates 
the situation of monopole-antimonople repulsion when a caloron dissociates 
into its constituents. 

The spatial coarse 
graining over (anti)selfdual calorons of charge modulus $|Q|=1$, which do not propagate (due to (anti)selfduality 
their energy-momentum tensor vanishes identically \cite{Shuryak}), 
yielding a highly accurate a priori estimate of the deconfining 
thermal ground state in terms of an inert, adjoint scalar field $\phi$ and a pure-gauge configuration $a^{\tiny\mbox{gs}}_\mu$, is performed over isolated and stable HS solutions \cite{Hofmann2012}. The coarse-grained field $a^{\tiny\mbox{gs}}_\mu$ represents a posteriori the effects of small holonomy changes 
due to (anti)caloron overlap and interaction.         

\subsection{Anatomy of a relevant Harrington-Shepard caloron\label{AHS}}

Let us now review \cite{GrossPisarskiYaffe1981} how the field strength of a HS caloron depends on the distance from its center at $\tau=r=0$. For $|x|\ll \beta$ ($|x|\equiv\sqrt{x^2}\equiv\sqrt{x_\mu x_\mu}$, $x_4\equiv\tau$) one has
\eqb
\label{prepotentialx<<beta}
\Pi(x)=(1+\frac{\pi}{3}\frac{s}{\beta})+\frac{\rho^2}{x^2}+O(x^2/\beta^2)\,,
\eqe
where $s$ is defined in Eq.\,(\ref{madistance}). From Eqs.\,(\ref{prepotentialx<<beta}) and 
(\ref{HScaloron}) one obtains with $|x|\ll \beta$ the following expression for $F_{\mu\nu}=\frac12\epsilon_{\mu\nu\kappa\lambda}F_{\kappa\lambda}\equiv\tilde{F}_{\mu\nu}$  
\eqb
\label{Finside}
F_{\mu\nu}^a=-4{\rho^\prime}^2 \frac{\bar{\eta}^a_{\alpha\beta}}{(x^2+{\rho^\prime}^2)^2} 
I_{\alpha\mu}I_{\beta\nu}+O(x^2/\beta^4)\,,
\eqe
where $I_{\alpha\mu}\equiv\delta_{\alpha\mu}-2\frac{x_\alpha x_\mu}{x^2}$. 
At small four-dimensional distances from the caloron center the field strength 
thus behaves like the one of a singular-gauge instanton with a 
renormalised scale parameter ${\rho^\prime}^2=\frac{\rho^2}{1+\frac{\pi}{3}\frac{s}{\beta}}$. Therefore, 
the field strength of the HS solution exhibits a dependence on $\tau$ and as such has no Minkowskian 
interpretation, see Sec.\,\ref{EMI}. What can be inferred for a Minkowskian spacetime though
is that the action of the configuration is attributable to winding of the caloron around the 
group manifold $S_3$ as induced by a spacetime point, the instanton center. This is because, in the sense 
of Eq.\,(\ref{actHS}), an instant has no analytic continuation or Wick rotation. 
(The 4D action or topological-charge {\sl density} of the caloron is regular at $\tau=r=0$, does 
depend on Euclidean spacetime in the vicinity of this point, and thus 
has no Minkowskian interpretation.)

For $r\gg\beta$ the selfdual electric and magnetic fields $E_i^a$ and $B_i^a$ are static and 
can be written as
\eqb
\label{EBlarge}
E^a_i=B_i^a\sim-\frac{\frac{\hat{x}^a\hat{x}_i}{r^2}-\frac{1}{rs}(\delta^a_i-3\hat{x}^a\hat{x}_i)}
{(1+\frac{r}{s})^2}\,,
\eqe
where $\hat{x}_i\equiv\frac{x_i}{r}$ and $\hat{x}^a\equiv\frac{x^a}{r}$. 
For $\beta\ll r\ll s$ Eq.\,(\ref{EBlarge}) simplifies as
\eqb
\label{EBlargeDyon}
E^a_i=B_i^a\sim-\frac{\hat{x}^a\hat{x}_i}{r^2}\,, 
\eqe
and thus describes a static non-Abelian monopole of unit electric and magnetic charges (dyon).  
For $r\gg s\gg\beta$ Eq.\,(\ref{EBlarge}) reduces to 
\eqb
\label{EBlargeDipole}
E^a_i=B_i^a\sim s\,\frac{\delta^a_i-3\,\hat{x}^a\hat{x}_i}{r^3}\,.
\eqe  
This is the field strength of a static, selfdual non-Abelian dipole field, its 
dipole moment $p_i^a$ given as
\eqb
\label{dipolemom}
p_i^a=s\,\delta_i^a\,.
\eqe
Interestingly, the same distance $s$, which sets the separation between the charge centers of an Abelian 
magnetic monopole and its antimonopole in a nontrivial-holonomy caloron, 
prescribes here for the case of trivial holonomy how small $r$ needs to be in order to reduce the non-Abelian 
dipole of Eq.\,(\ref{EBlargeDipole}) to the non-Abelian monopole constituent, 
see Eq.\,(\ref{EBlargeDyon}). For a HS anticaloron one simply replaces $E^a_i=B_i^a$ by $E^a_i=-B_i^a$ 
in Eqs.\,(\ref{EBlarge}), (\ref{EBlargeDyon}), and (\ref{EBlargeDipole}). 

Finally, let us remark that the condition $s\gg\beta$, which is required for Eqs.\,(\ref{EBlargeDyon}) and (\ref{EBlargeDipole}) to be valid, is always satisfied for the caloron scale 
$\rho\sim |\phi|^{-1}$ which is relevant for the building of the thermal 
ground state in the deconfining phase of SU(2) Yang-Mills thermodynamics \cite{Hofmann2012}. Namely, 
one has
\eqb
\label{rhorequ}
\frac{s}{\beta}=\pi\left(\frac{\rho}{\beta}\right)^2=
\pi\left(\frac{\lambda^{3/2}}{2\pi}\right)^2=\frac{\lambda^{3}}{4\pi}\ge 212.3\,,
\eqe
where $\lambda\equiv\frac{2\pi T}{\Lambda}\ge\lambda_c=13.87$.

\section{Thermal ground state as induced by a probe\label{EMvacTGS}}

The postulate that thermal photon propagation should be described by an SU(2) rather 
than a U(1) gauge principle was put forward in \cite{Hofmann2005} and 
has undergone various levels of investigation ever since, see 
\cite{Hofmann2012,NaturePhys2013,Hofmann2015}. As a result, 
the associated Yang-Mills 
scale $\Lambda\sim 1.0638\times 10^{-4}\,$eV is fixed by low-frequency 
observation of the Cosmic Microwave Background (CMB) \cite{Arcade2} to correspond 
to the critical temperature for the deconfining-preconfining phase transition 
being the CMB's present baseline temperature $T_0=2.725\,$K \cite{Hofmann2009}. This prompted 
the name SU(2)$_{\tiny\mbox{CMB}}$. In the following we would like to 
investigate in what sense the vacuum parameters of classical 
electrodynamics, namely the electric permittivity $\epsilon_0$ and the 
magnetic permeability $\mu_0$, can be reduced to the physics of the static, 
non-Abelian, and (anti)selfdual monopole and dipole configurations represented by 
HS (anti)calorons in the regimes $\beta\ll r\ll s$ and $r\gg s\gg\beta$, respectively, 
see Sec.\,\ref{AHS}. To do this, the concept of a thermal ground state 
together with information on how it is obtained \cite{Hofmann2012} 
as well as the results of Sec.\,\ref{AHS} \cite{GrossPisarskiYaffe1981} are invoked.

\subsection{Preexisting dipole densities\label{PDD}}

Let us discuss the case $r\gg s$. In order to 
not affect spatial homogeneity on scales comparable to 
or smaller than $s$ the 
electromagnetic field, which propagates through the deconfining 
thermal ground state in the absence of any explicit electric charges, is considered a 
plane wave of wave 
length $l$ much larger than $s$. Such a field effectively sees a density of selfdual 
dipoles, see Eq.\,(\ref{EBlargeDipole}). Because they are given by $p_i^a=s\delta^a_i$ 
their dipole moments align along the direction of the exciting electric or magnetic field both in 
space and in the SU(2) algebra su(2). Note that at this stage the definition of what is to be viewed as an 
Abelian direction in su(2) is a global gauge convention such that {\sl all} spatial 
directions of the dipole moment $p_i^a$ are a priori thinkable. That is, 
dynamical Abelian projection of the non-Abelian situation 
of Eq.\,(\ref{EBlargeDipole}) is owed to the Abelian and dipole aligning nature of the exciting, massless 
field \cite{Hofmann2012}. Modulo global gauge rotations, this field is exists 
because of the adjoint Higgs mechanism invoked by the inert field $\phi$. 

Per spatial coarse-graining volume $V_{\tiny\mbox{cg}}$ of radius $|\phi|^{-1}=\rho=\sqrt{\frac{\Lambda^3}{2\pi T}}$ with 
\eqb
\label{cgvol}
V_{\tiny\mbox{cg}}=\frac43\pi|\phi|^{-3}\,,
\eqe
the center of a selfdual HS caloron 
and the center of an antiselfdual HS anticaloron \cite{Hofmann2012} reside. Note the large hierachy between $s$ 
(the minimal spatial distance to the center of a (anti)caloron, which allows to identify 
the static, (anti)selfdual dipole) and the radius of the sphere $|\phi|^{-1}$ defining 
$V_{\tiny\mbox{cg}}$, 
\eqb
\label{ratiophim1s}
\frac{s}{|\phi|^{-1}}=\frac12\lambda^{3/2}\ge 25.83\,\left(\frac{\lambda}{\lambda_c}\right)^{3/2}\,.
\eqe
If the exciting field is electric then it 
sees {\sl twice} the electric dipole $p_i^a$ (cancellation of magnetic dipole between caloron and anticaloron), if it is magnetic it sees {\sl twice} the magnetic dipole $p_i^a$ (cancellation of electric dipole between caloron and anticaloron, $\vec{E}=-\vec{B}\ \ \Leftrightarrow\ \ -\vec{E}=\vec{B}$). 
To be definite, let us discuss the 
electric case in detail, characterised by an 
exciting Abelian field $\vec{E}_e$. 
The modulus of the according dipole density $\vec{D}_e||\vec{E}_e$ is given as  
\eqb
\label{dipoledensity}
|\vec{D}_e|=\frac{2s}{V_{\tiny\mbox{cg}}}=
\frac{3}{4\pi}\Lambda^2\lambda^{1/2}_c\left(\frac{\lambda}{\lambda_c}\right)^{1/2}\,.
\eqe 
In classical electromagnetism the relation between the fields $\vec{E}_e$ and $\vec{D}_e$ is 
\eqb
\label{permvac}
\vec{D}_e=\epsilon_0\vec{E}_e\,,
\eqe
where 
\eqb
\label{measuredeps}
\epsilon_0=5.52703\times 10^7\,\frac{Q}{\mbox{V\,m}}
\eqe 
is the electric 
permittivity of the vacuum, and $Q=1.602\times 10^{-19}\,$A\,s denotes the 
electron charge (unit of elementary charge), now both in SI units.

According to electromagnetism the energy density $\rho_{\tiny\mbox{EM}}$ carried by 
an external electromagnetic wave with $|\vec{E}_e|=|\vec{B}_e|$ is
\eqb
\label{eneE}
\rho_{\tiny\mbox{EM}}=\frac{1}{2}(\epsilon_0\vec{E}_e^2+\frac{1}{\mu_0}\vec{B}_e^2)
=\frac{1}{2}(\epsilon_0+\frac{1}{\mu_0})\vec{E}_e^2\,.
\eqe
In natural units we have $\epsilon_0\mu_0=1/c^2=1$, and therefore\footnote{To assume $\epsilon_0\mu_0=1$ 
just represents a short cut, it would have come out automatically if we had 
treated the magnetic case explicitly.} $\mu_0=1/\epsilon_0$. 
Thus
\eqb
\label{eneready}
\rho_{\tiny\mbox{EM}}=\epsilon_0\vec{E}_e^2\,.
\eqe
The $\vec{E}_e$-field dependence of $\rho_{\tiny\mbox{EM}}$ is converted into a 
fictitious temperature dependence by demanding that the temperature of the 
thermal ground state of SU(2)$_{\tiny\mbox{CMB}}$ adjusts itself 
such as to accomodate $\rho_{\tiny\mbox{EM}}$,
\eqb
\label{TE}
\rho_{\tiny\mbox{EM}}=4\pi\Lambda^3 T\ \ \ \Leftrightarrow\ \ \ \ |\vec{E}_e|=\Lambda^2\sqrt{2\frac{\lambda_c}{\epsilon_0}}\left(\frac{\lambda}{\lambda_c}\right)^{1/2}\,. 
\eqe
Eq.\,(\ref{TE}) generalises 
the thermal situation of ground-state energy density of Sec.\,\ref{RGF}, 
where ground-state thermalisation is induced by a thermal ensemble 
of excitations, to the case where the thermal ensemble is missing but 
the probe field induces a fictitious temperature and energy density 
to the ground state. Combining Eqs.\,(\ref{dipoledensity}), (\ref{permvac}), 
and (\ref{TE}), and introducing the ratio $\xi$ 
 between the non-Abelian monopole charge $Q^\prime$ in the dipole and the (Abelian) 
 electron charge\footnote{In natural units, the actual charge of the 
monopole constituents within the (anti)selfdual dipole is $1/g$ where $g$ is the 
undetermined fundamental gauge coupling. This is absorbed into $\xi$.} 
$Q$, we obtain
\eab
\label{eps0}
\epsilon_0[Q\mbox{(V\,m)}^{-1}]&=&
\frac{3}{\sqrt{32}\pi}\left(\frac{\Lambda[\mbox{m}^{-1}]}{\Lambda[\mbox{eV}]}\right)^{1/2}\xi\sqrt{\epsilon_0[Q\mbox{(V\,m)}^{-1}]}\ \ \ 
\Leftrightarrow \nonumber\\ 
\epsilon_0[Q\mbox{(V\,m)}^{-1}]&=&\frac{9}{32\pi^2}\frac{\Lambda[\mbox{m}^{-1}]}{\Lambda[\mbox{eV}]}\,
\xi^2\,.
\eae 
Notice that $\epsilon_0$ does not exhibit any temperature dependence and thus no dependence 
on the field strength $\vec{E}_e$. It is a universal constant. 
In particular, $\epsilon_0$ does {\sl not} relate to the state of fictitious 
ground-state thermalisation which would associate to the rest frame of a local heat bath.

To produce the measured value for $\epsilon_0$ as in Eq.\,(\ref{measuredeps}) the ratio $\xi$ in 
Eq.\,(\ref{eps0}) is required to be
\eqb
\label{xi}
\xi\equiv\frac{Q^\prime}{Q}=19.56\,.
\eqe 
Thus, compared to the electron charge, the charge unit associated with a (anti)selfdual non-Abelian dipole, 
residing in the thermal ground state, is gigantic.   

Discussing $\mu_0$, we could have been proceeded in complete analogy to the case of $\epsilon_0$.
(It would be $\mu_0^{-1}$ defining the ratio between 
the modulus of the magnetic dipole density and the magnetic flux density $|\vec{B}|$.) 
Here, however, the comparison between non-Abelian magnetic charge and 
an elementary, magnetic, and Abelian charge is not facilitated since the latter does not 
exist in electrodynamics.

Finally, let us see what the condition that the wavelength $l$ of the 
electromagnetic disturbance considered in this section is much larger than $s$ 
means in units of meters when invoking SU(2)$_{\tiny\mbox{CMB}}$. One has 
\eqb
\label{lambdaggs}
l\gg \frac{\lambda_c^2}{2\Lambda}\left(\frac{\lambda}{\lambda_c}\right)^2=
1.1254\,\mbox{m}\,\left(\frac{T}{2.725\,\mbox{K}}\right)^2\,.   
\eqe
Setting $T=T_c=2.725\,$K in Eq.\,(\ref{lambdaggs}), 
we obtain a lower bound on the wave length of $l_{\tiny\mbox{min}}=1.1254\,\mbox{m}$.    
 
\subsection{Explicitly induced dipole densities\label{IDD}} 

Let us now discuss the case $\beta\ll |\phi|^{-1}\ll r\ll s$. To rely on the presence of the 
inert adjoint scalar field $\phi$ of the effective theory, $r$ needs to be 
larger than the spatial coarse-graining scale 
$|\phi|^{-1}=\frac{1}{2\pi}\lambda^{3/2}_c\left(\frac{\lambda}{\lambda_c}\right)^{3/2}\beta\ge 8.22\,\beta$. Within 
the according regime $|\phi|^{-1}\le r\ll s$ 
of spatial distances from the caloron center at $(\tau=0,\vec{x}=0$) an 
electromagnetic wave of wave length $l$ sees the 
selfdual field of a static, non-Abelian monopole of 
electric and magnetic charge as in Eq.\,(\ref{EBlargeDyon}) 
which is centered at $\vec{x}=0$. A selfdual Abelian field strength $E_i=B_i$ of this  
monopole is obtained \cite{GoddardOlive} as
\eqb
\label{gaugetransmon}
E_i=B_i=\frac{\phi^a}{|\phi|} E_i^a=\frac{\phi^a}{|\phi|} B_i^a
\eqe
with the field $\phi$ gauged from unitary gauge $\phi^a=2|\phi|\delta^{a3}$ 
into "hedgehog" gauge $\phi^a=2|\phi|\hat{x}^a$. 
The according gauge transformation is give in terms of the group element 
$\Omega\equiv \cos\frac12\psi-i\hat{k}\cdot\vec{\sigma}\sin\frac12\psi$ where $\sigma_i\,, (i=1,2,3)\,,$ 
are the Pauli matrices, $\hat{k}\equiv\frac{\hat{e}_3\times\hat{x}}{\sin\theta}$, $\hat{e}_3$ is the 
third vector of an orthonormal basis of space, $\theta\equiv\angle(\hat{e}_3,\hat{x})$, 
and $\psi=\theta$ for $0\le\theta\le\pi-\epsilon$, which 
smoothly drops to zero at $\theta=\pi$, and the limit $\epsilon\to 0$ 
is understood \cite{GoddardOlive}. For the monopole field $E_i$ 
to be normalized to charge $-2Q^\prime$ one\footnote{The factor two in front of the monopole 
charge $Q^\prime$ is due to a contribution to the monopole field strength 
of the anticaloron identical to that of the caloron.} thus has   
\eqb
\label{Abelianmon}
E_i=B_i=-\frac{2Q^\prime}{4\pi\epsilon_0}\frac{\hat{x}_i}{r^2}=
-\frac{2Q^\prime\mu_0}{4\pi}\frac{\hat{x}_i}{r^2}\,.
\eqe
The electric or magnetic poles of Eq.\,(\ref{Abelianmon}) should 
independently react by harmonic and linear 
acceleration to the presence of an external electric or magnetic field $\vec{E}_{e}$ or $\vec{B}_{e}$, respectively, forming a monochromatic electromagnetic wave of frequency $\omega=\frac{2\pi}{l}$. 
At $\vec{x}=0$ one has 
\eqb
\label{harmonic}
\vec{E}_{e}=\vec{E}_0\,\sin(\omega t)\,,
\eqe
and readily derives (as in Thomson scattering) that the induced dipole 
moment $\vec{p}$, say, for the electric case, is given as
\eqb
\label{dipoleinduced}
\vec{p}=-\frac{\vec{E}_e (2\,Q^\prime)^2}{m\omega^2}\,.
\eqe
Interestingly, by virtue of Eq.\,(\ref{Abelianmon}) the squared charge of the pole, 
$(2\,Q^\prime)^2$, cancels out in $\vec{p}$ because its mass $m$ carries an identical factor (only the 
electric (magnetic) monopole is linearly and harmonically accelerated by the external electric (magnetic) 
field $\vec{E}_{e}$ ($\vec{B}_{e}$) and hence $m$ carries electric (magnetic) field energy only): 
\eab
\label{mass}
m&=&\frac12\epsilon_0\,4\pi\int_{|\phi|^{-1}}^\infty dr\,r^2\,E_i E_i\nonumber\\ 
&=&\frac{1}{8\pi\epsilon_0}(2\,Q^\prime)^2\int_{|\phi|^{-1}}^\infty \frac{dr}{r^2}=
\frac{1}{8\pi\epsilon_0}(2\,Q^\prime)^2 |\phi|\ \ \ \Rightarrow\nonumber\\ 
\vec{p}&=&-\frac{8\pi\epsilon_0\vec{E}_e}{|\phi|\omega^2}\,. 
\eae
Again, the volume $V_{\tiny\mbox{cg}}$, which underlies the dipole 
moment $\vec{p}$ by containing a caloron and an 
anticaloron center, is given by Eq.\,(\ref{cgvol}), and we have
\eqb
\label{Dmon}
|\vec{D}_e|=\frac{|\vec{p}|}{V_{\tiny\mbox{cg}}}=6\,\epsilon_0\frac{|\vec{E}_e||\phi|^2}{\omega^2}\,,
\eqe
and therefore 
\eqb
\label{epsilonmon}
\epsilon_0\equiv\frac{|\vec{D}_e|}{|\vec{E}_e|}=
6\,\epsilon_0\frac{|\phi|^2}{\omega^2}\,.
\eqe
In Eq.\,(\ref{epsilonmon}) also the vacuum permittivity $\epsilon_0$ 
cancels out, and we are left with the condition 
\eqb
\label{condonomega}
\omega=\sqrt{6}\,|\phi|\ \ \ \Leftrightarrow \ \ \ 
l=\sqrt{\frac{2}{3}}\pi |\phi|^{-1}=\sqrt{\frac{2}{3}}\pi\Lambda^{-1}\lambda^{1/2}_c
\left(\frac{\lambda}{\lambda_c}\right)^{1/2}\,,
\eqe 
where temperature $T$ (or $\lambda$), again, is set by the local field 
strengths of the electromagnetic probe according to Eqs.\,(\ref{eneE}) 
and (\ref{TE}). Let us see whether the second of Eqs.\,(\ref{condonomega}) is consistent 
with  $|\phi|^{-1}\le r=l\ll s$. The former inequality is selfevident, and 
the latter follows from
\eqb
\label{soverl}
\frac{s}{l}=\sqrt{\frac{3}{8}}\frac{\lambda_c^{3/2}}{\pi}\left(\frac{\lambda}{\lambda_c}\right)^{3/2}=
10.069\,\left(\frac{\lambda}{\lambda_c}\right)^{3/2}\,.
\eqe 
By setting $\lambda=\lambda_c$ we obtain from 
Eqs.\,(\ref{condonomega}) a minimal wavelength
\eqb
\label{minwl}
l_{\tiny\mbox{min}}=\sqrt{\frac{2}{3}}\pi\Lambda^{-1}\lambda^{1/2}_c=0.112\,\mbox{m}\,.
\eqe
This wavelength is about a factor of ten smaller than the lowest possible value as expressed by 
Eq.\,(\ref{lambdaggs}). 

\subsection{Discussion}

In Secs.\,\ref{PDD} and \ref{IDD} an analysis 
was performed to clarify to what extent the thermal ground state 
of SU(2)$_{\tiny\mbox{CMB}}$ can be regarded as the luminiferous aether, 
supporting the propagation of an external electromagnetic wave (probe) of field strengths 
$|\vec{E}_e|=|\vec{B}_e|$ and wave length $l$ which, by itself, is not thermal. 

Sec.\,\ref{PDD} has focussed on wave lengths that are large compared to the 
distance $s=\frac{\pi|\phi|^{-2}}{\beta}$, very large compared to the resolution 
limit $|\phi|^{-1}$ of the effective theory for 
deconfining SU(2)$_{\tiny\mbox{CMB}}$ and even more so on the scale 
of inverse temperature $\beta$, see Eq.\,(\ref{rhorequ}), when 
(anti)calorons of SU(2)$_{\tiny\mbox{CMB}}$ manifest themselves as static 
(anti)selfdual dipoles whose dipole moment is set by a 
fictitious temperature representing the intensity of the probe via 
Eq.\,(\ref{TE}). And indeed, in this case vacuum permittivity $\epsilon_0$ 
and permeability $\mu_0$ turn out to be universal constants, see Eq.\,(\ref{eps0}). When 
confronted with their experimental values the charges of 
the "constituent" non-Abelian monopoles in a dipole follow in units of 
electron charge, see Eq.\,(\ref{xi}).               

Eqs.\,(\ref{lambdaggs}) and (\ref{TE}) indicate that an uncertainty-like relation between field $|\vec{E}_e|$ 
strength and wave length $l$ takes place as follows  
\eqb
\label{intwavelength}
|\vec{E}_e|^4 l^{-1}\ll\frac{8\Lambda^9}{\epsilon^2_0}\,.
\eqe
Therefore, the larger the probe intensity the longer its wave 
length is required to be in order to be supported by thermal ground-state 
physics. In any case, in SU(2)$_{\tiny\mbox{CMB}}$ wave lengths need to be larger than the meter 
scale, see Eq.\,(\ref{lambdaggs}).  

Things are different for wave lengths that are large on the scale 
$|\phi|^{-1}$ but short on the scale $s=\frac{\pi|\phi|^{-2}}{\beta}$. This case is 
investigated in Sec.\,\ref{IDD}. 
Then a (anti)caloron can no longer be viewed as a static, (anti)selfdual dipole but 
rather is represented by a static, (anti)selfdual monopole. 
However, an attempt to consider dipole moments as induced dynamically by monopole shaking through the probe fields 
renders the definitions of vacuum parameters 
$\epsilon_0$ and $\mu_0$ meaningless, see Eq.\,(\ref{epsilonmon}). It does yield a 
fixation of the probe's wave length $l$ in terms of $|\phi|^{-1}$ though, 
see Eq.\,(\ref{condonomega}). While the former situation is not 
surprising because single magnetic charges violate the 
Bianchi identities for the electromagnetic field strength  
tensor $F_{\mu\nu}$ it is nontrivial that $l$ turns 
out to selfconsistently satisfy the constraint that 
$s\gg l>|\phi|^{-1}$. Note that the minimal wave lengths $l_{\tiny\mbox{min}}=1.1254\,\mbox{m}$ and $l_{\tiny\mbox{min}}=\sqrt{\frac{2}{3}}\pi\Lambda^{-1}\lambda^{1/2}_c=0.112\,\mbox{m}$ as 
obtained in Secs.\,\ref{PDD} and \ref{IDD}, respectively, are off by a factor of ten only.

\section{Summary and Conclusions\label{SC}}

We have addressed the question how the concept of a thermal 
ground state of SU(2)$_{\tiny\mbox{CMB}}$, 
which in a fully thermalised situation coexists 
with a spectrum of partially massive (adjoint Higgs mechanism) thermal 
excitations of the same temperature, can 
be employed to understand the propagation of a nonthermal probe 
(monochromatic electromagnetic wave) in vacuum, characterised 
by electric permittivity $\epsilon_0$ and magnetic permeability $\mu_0$. 
To do this, we have appealed to the fact 
that the thermal ground state emerges by a spatial coarse 
graining over (anti)selfdual fundamental Yang-Mills 
fields of topological charge modulus unity at finite temperature: Harrington-Shepard 
(anti)calorons of trivial holonomy. Note that this coarse-graining does not require the considerations of thermal 
excitations. Therefore, it is suggestive that the concept of the thermal ground state can be extended 
to the description of a nonthermal situation with the parameter $T$ acting as 
a period in compactified Euclidean spacetime $S_1\times{\bf R}^3$ and no longer as 
a thermodynamical temperature.     

Knowing how large the 
coarse-graining volume is, which contains one caloron and one anticaloron center, where 
the unit of action $\hbar$ is localised (Sec.\,\ref{RGF}),  
and by exploiting the structure of these field configurations spatially 
far away (Sec.\,\ref{AHS}) from their centers, we were able to deduce densities of 
electric and magnetic dipoles in Sec.\,\ref{PDD}. Dividing these dipole densities by 
the respective field strengths of the probe, 
selfconsistently adjusted to the energy-density of the 
thermal ground state (small, transient (anti) caloron holonomies), 
yields definitions of $\epsilon_0$ and $\mu_0$. In the electric case a match with the 
experimental value predicts the charge of one of the monopoles, which 
constitutes the dipole, in terms of electron charge. The former charge turns out to be substantially 
larger than the latter. 

As shown in Sec.\,\ref{IDD} this way of reasoning, which is valid for large wave 
lengths ($l\gg s$) only, cannot 
be extended to smaller wave lengths $l$. Namely, in a region of spatial distances to the 
(anti)caloron center, where the configuration resembles (anti)selfdual, static 
monopoles, the definition of $\epsilon_0$ and $\mu_0$ in terms of 
dipole densities that are explicitly induced by the probes oscillating 
field strengths becomes meaningless. This is expected since the existence 
of resolved magnetic monopoles would violate the Bianchi 
identities for the field strength tensor $F_{\mu\nu}$ 
of electromagnetism. 
    
We conclude that the thermal ground state of 
SU(2)$_{\tiny\mbox{CMB}}$ supports the propagation 
of a nonthermal probe purely in terms of Harrington-Shepard 
(anti)calorons (trivial holonomy) if the probe's wave length $l$ is sufficiently 
large (the regime 
$l\gg s=\frac{\pi|\phi|^{-2}}{\beta}\ge 1.1254\,$m) and that there  
is an uncertainty-like relation between $l$ and the square of the probe's 
intensity, see (\ref{intwavelength}).  

To address the nonthermal propagation of shorter wavelength and/or higher intensities, see Eq.\,(\ref{intwavelength}), additional, mixing SU(2) gauge factors of hierarchically larger Yang-Mills scales have to be 
postulated, see discussion in \cite{Hofmann2015}. At present, however,  
it is not clear how the effectiveness of the very successful 
Standard Model of particle physics in describing electroweak processes can be achieved in terms of 
such a more fundamental framework of pure Yang-Mills dynamics.

\section*{Acknowledgments}   

One of us (RH) would like to thank Stan Brodsky for a stimulating conversation.

\end{document}